\documentclass{iau} 
\usepackage{graphicx}

\title[The VISCACHA survey] 
{The VISCACHA survey -- deep and resolved photometry of star clusters in the Magellanic Clouds}

\author[Bruno Dias and the VISCACHA team]   
{Bruno Dias$^{1,2}$,
Francisco Maia$^{3}$,
Leandro Kerber$^{4}$,
João F.C. dos Santos Jr.$^{5}$,
Eduardo Bica$^{6}$,
Tina Armond$^{7}$,
Beatriz Barbuy$^{8}$,
Luciano Fraga$^{9}$,
Jose A. Hernandez-Jimenez$^{2}$,
Orlando J. Katime Santrich$^{4}$,
Raphael A. P. Oliveira$^{8}$,
Angeles P\'erez-Villegas$^{8}$,
Andres Piatti$^{10,11}$,
Bruno Quint$^{12}$,
David Sanmartin$^{12}$,
Mateus S. Angelo$^{13}$,
Stefano O. Souza$^{8}$,
Rodrigo G. Vieira$^{8}$,
Pieter Westera$^{14}$,
Celeste Parisi$^{10,15}$,
Doug Geisler$^{16,17,18}$,
Dante Minniti$^{2,19,20}$,
Roberto Saito$^{21}$,
Lilia Bassino$^{22,10}$,
Bruno De Bortoli$^{22,10}$,
André Figueiredo$^{8}$,
\and Leandro Rímulo$^{23}$
}

\affiliation{
$^{1}$European Southern Observatory, Alonso de C\'ordova 3107, Vitacura 19001, Chile \\ 
$^{2}$Departamento de F\'{\i}sica, Facultad de Ciencias Exactas, Universidad Andr\'es Bello, Av. Fernandez Concha 700, Las Condes, Santiago, Chile\\  
$^{3}$Instituto de F\'isica - UFRJ, Bloco A Centro de Tecnologia, Av. Athos da Silveira Ramos, 149 - Cidade Universit\'aria, Rio de Janeiro, 21941-909, Brazil \\  
$^{4}$Universidade Estadual de Santa Cruz, Depto. de Ci\^encias Exatas e Tecnol\'ogicas, Rodovia Jorge Amado km 16, 45662-900, Brazil \\  
$^{5}$Universidade Federal de Minas Gerais, ICEx, Av. Ant\^onio Carlos 6627, 31270-901, Brazil \\  
$^{6}$Universidade Federal do Rio Grande do Sul, Instituto de F\'isica, Av. Bento Gon\c calves 9500, 91501-970 ,Brazil \\   
$^{7}$Universidade Federal de S\~ao Jo\~ao del-Rei, Departamento de Estat\'istica, F\'isica e Matem\'atica, Campus Alto Paraopeba, Rod.: MG 443, KM 7, Ouro Branco - MG, 36420-000, Brazil \\
$^{8}$Universidade de S\~ao Paulo, IAG, Rua do Mat\~ao 1226, 05508-090, Brazil \\  
$^{9}$Laborat\'orio Nacional de Astrof\'isica, Rua Estados Unidos 154, 37504-364, Brazil\\  
$^{10}$Consejo Nacional de Investigaciones  Cient\'{\i}ficas y T\'ecnicas, Av. Rivadavia 1917, C1033AAJ, Buenos Aires, Argentina\\  
$^{11}$Observatorio Astron\'omico de C\'ordoba, Laprida 854, 5000, C\'ordoba, Argentina\\  
$^{12}$Gemini Observatory, c/o AURA - Casilla 603, La Serena, Chile\\  
$^{13}$Centro Federal de Educa\c c\~ao Tecnol\'ogica de Minas Gerais, Av. Monsenhor Luiz de Gonzaga, 103, Nepomuceno 37250-000, Brazil\\  
$^{14}$Universidade Federal do ABC, Centro de Ci\^encias Naturais e Humanas, Avenida dos Estados, 5001, 09210-580, Brazil\\  
$^{15}$Instituto de Astronom\'ia Te\'orica y Experimental, Laprida 854, C\'ordoba, Argentina\\  
$^{16}$Departamento de F\'isica y Astronom\'ia, Universidad de La Serena, Avenida Juan Cisternas 1200, La Serena, Chile \\  
$^{17}$Instituto de Investigaci\'on Multidisciplinario en Ciencia y Tecnolog\'ia, Universidad de La Serena. Benavente 980, La Serena, Chile \\  
$^{18}$Departmento de Astronom\'ia, Universidad de Concepci\'on, Casilla 160-C, Concepci\'on, Chile \\  
$^{19}$Millennium Institute of Astrophysics, Av. Vicu\~na Mackenna 4860, 782-0436 Macul, Santiago, Chile\\  
$^{20}$Vatican Observatory, Vatican City State V-00120, Italy\\  
$^{21}$Departamento de F\'isica, Universidade Federal de Santa Catarina, Trindade 88040-900, Florian\'opolis, SC, Brazil\\  
$^{22}$Facultad de Ciencias Astron\'omicas y Geof\'isicas de la Universidad Nacional de La Plata, and Instituto de Astrof\'isica de La Plata (CCT La Plata - CONICET, UNLP), Paseo del Bosque S/N, B1900FWA La Plata, Argentina\\  
$^{23}$Universidad de los Andes, Departamento de F\'isica, Carrera 1 18A-10, Bloque Ip. Bogot\'a - Colombia\\  
}

\pubyear{2019}
\volume{351}  
\jname{Star Clusters: From the Milky Way to the Early Universe}
\editors{A. Bragaglia, M.B. Davies, A. Sills \& E. Vesperini, eds.}
\begin{document}

\maketitle

\begin{abstract}
The VISCACHA (VIsible Soar photometry of star Clusters in tApii and Coxi HuguA\footnote{LMC and SMC names in the Tupi-Guarani language spoken by native people in Brazil}) Survey is an ongoing project based on deep and spatially resolved photometric observations of Magellanic Cloud star clusters, collected using the SOuthern Astrophysical Research (SOAR) telescope together with the SOAR Adaptive Module Imager. So far we have used $>$300h of telescope time to observe $\sim$150 star clusters, mostly with low mass ($M < 10^4 M_{\odot}$) on the outskirts of the LMC and SMC. With this high-quality data set, we homogeneously determine physical properties using deep colour-magnitude diagrams (ages, metallicities, reddening, distances, mass, luminosity and mass functions) and structural parameters (radial density profiles, sizes) for these clusters which are used as a proxy to investigate the interplay between the Magellanic Clouds and their evolution. We present the VISCACHA survey and its initial results, based on our first two papers. The project's long term goals and expected legacy to the community are also addressed.
\end{abstract}

\firstsection 
\section{Introduction}

\begin{figure}[htb]
   \centering
   \includegraphics[width=0.8\columnwidth]{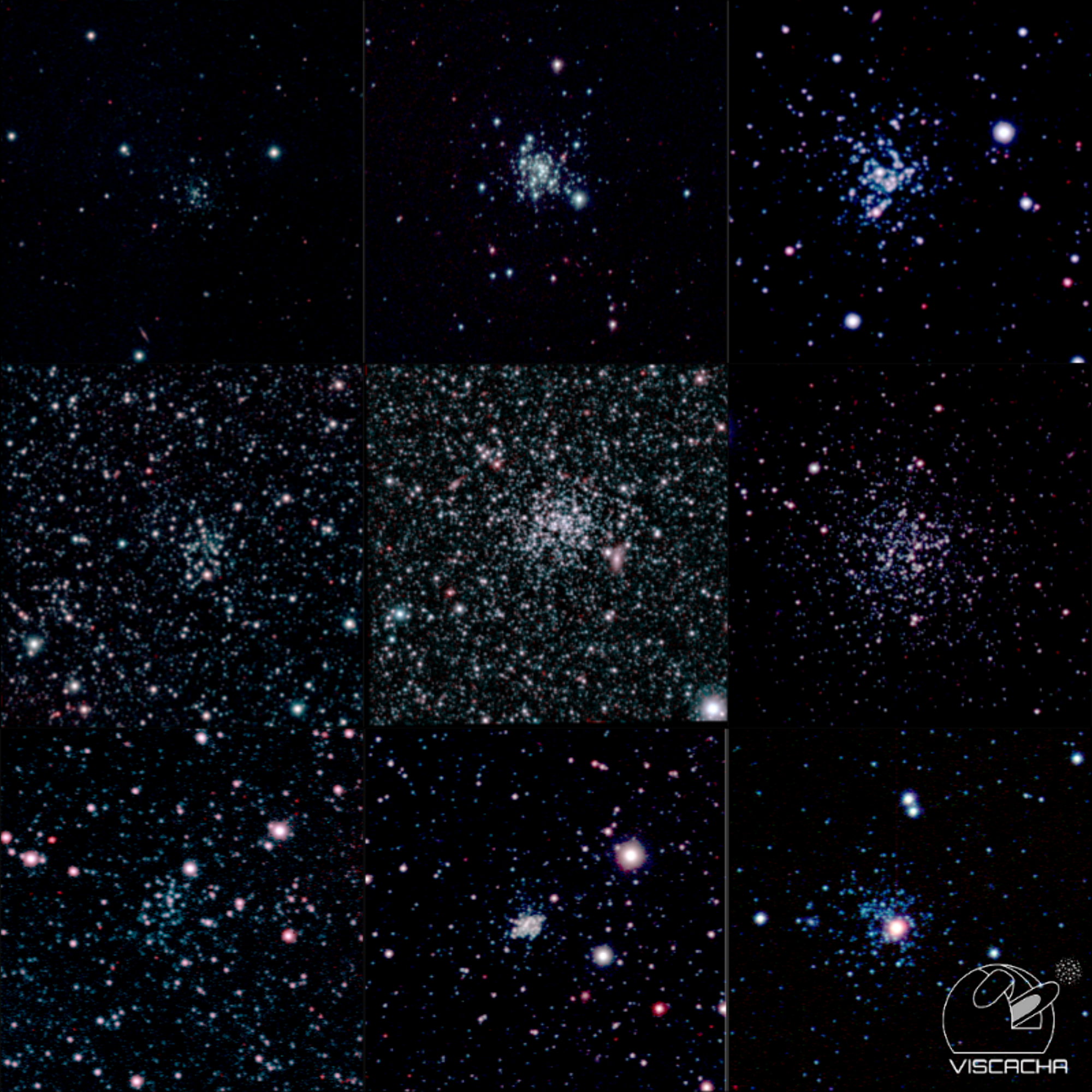}
   \caption{Mosaic of coloured images of the nine clusters analysed in Paper I (Maia et al. 2019) based on V,I images. Each image has a size of $3'\times3'$, North is up, East is left.}
    \label{fig1}%
    \end{figure}

\begin{figure}[htb]
   \centering
   \includegraphics[width=0.75\columnwidth]{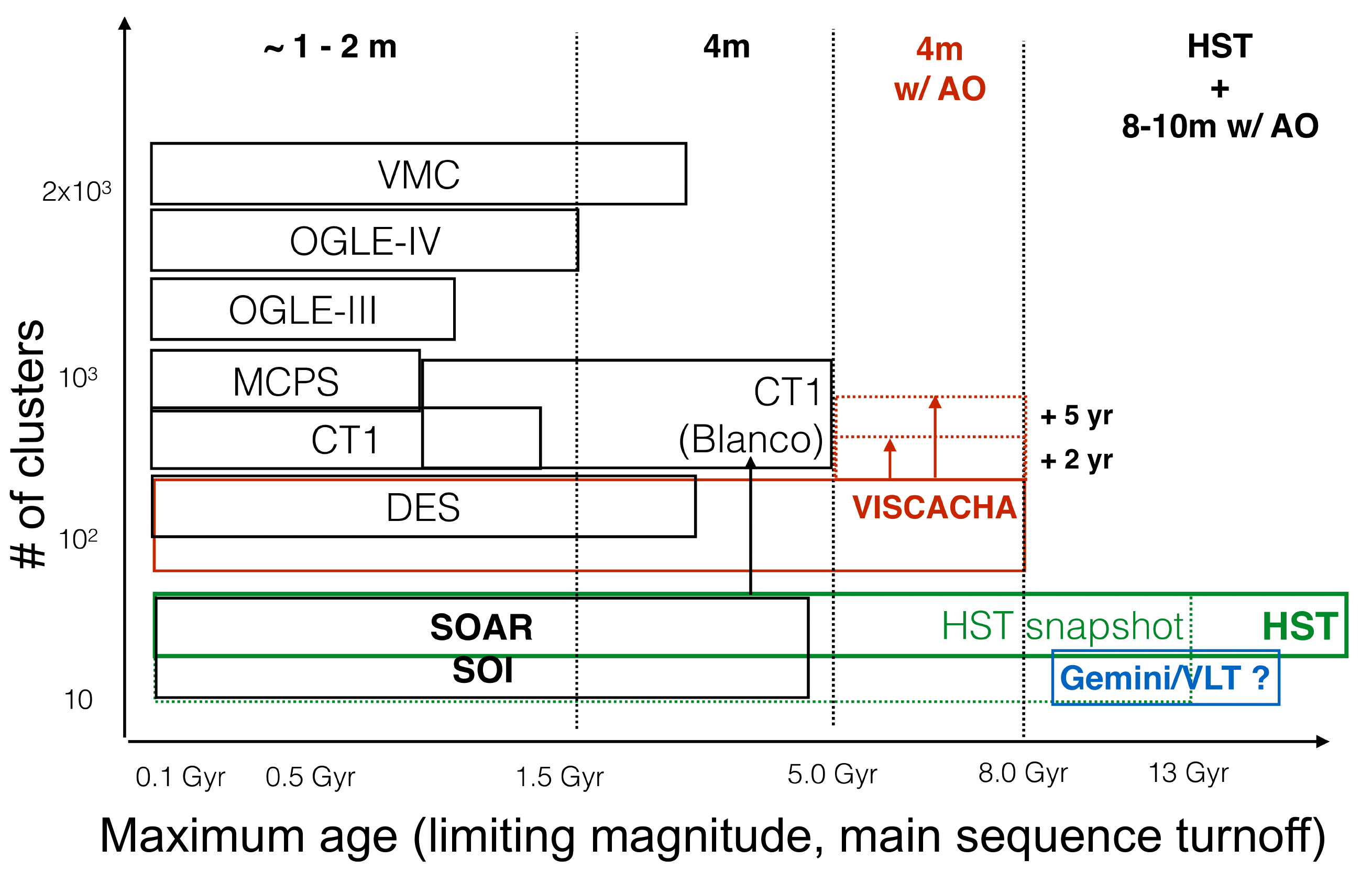}
   \caption{Star cluster age limit (main sequence turnoff magnitude limit) reached by different photometric surveys and telescopes versus the number of clusters. Usually there is a compromise to either observe a large area on the sky with shallow photometry or a deep photometry for a few objects. One limitation is the image quality, which the VISCACHA survey overcomes using the 4m SOAR telescope with AO reaching FWHM = 0.4" and 0.5" in I and V filters.}
    \label{fig2}%
    \end{figure}

The VISCACHA project (PI: B.Dias, paper I: Maia et al. 2019) is a deep photometric survey observing
star clusters in the Large and Small Magellanic Clouds (LMC, SMC). In contrast with other large area sky surveys
pointing at the Magellanic system --- such as the VMC (Cioni et al. 2011) or the SMASH (Nidever et al. 2017) --- our 
survey is focused on star clusters throughout the SMC, LMC, and Bridge.
Our advantage is the goal of obtaining a final image quality a factor 2-3 better than the large surveys, improving the spatial resolution (see Figs. \ref{fig1} and \ref{fig2}). 
We also obtain accurate and deep photometry reaching a few magnitudes fainter than the larger surveys in the crowded regions such as star clusters. 
HST, Gemini, VLT can reach older clusters, but usually focus on a few more massive objects. The niche explored by VISCACHA is unique. i.e., obtaining deep and resolved photometry for a relatively large number of star clusters (Fig. \ref{fig2}).
The strategy is that we also use a 4m class telescope as most of the modern surveys, but we
combine with the use of adaptive optics (AO) that makes the VISCACHA data distinct and complementary 
to any other previous photometric survey on the Magellanic Clouds. 

The SOuthern Astrophysical Research (SOAR) telescope is accessible to our team through Brazil and Chile hosting 
together 41\% of the available nights. We are using Brazilian and Chilean time to accomplish our planned
observations using the SOAR Adaptive Module Imager (SAMI) since its commissioning in 2015. Some of the 
members are from Argentina, that also has access to Gemini telescope, and we are using joint Brazilian, Chilean, 
and Argentinean time for spectroscopic follow-up observations with GMOS on Gemini South.

Among the main topics that we address in the VISCACHA project we list (i) 3D distribution of the SMC and LMC star clusters;
(ii) star formation history and chemical evolution history per region; (iii) formation and dissolution history of star clusters in the Magellanic Clouds;
(iv) dependence of structural parameters with spatial distribution and age; (v) combination with radial velocities and proper motions to derive orbits and timescales of
interactions between the galaxies; (vi) initial mass function for clusters of different total masses, among others. We present here an overview of a few projects we have been developing within the collaboration.

\section{The VISCACHA projects}


We have developed sophisticated tools to estimate the cluster membership probability for each star and to fit the observed colour-magnitude diagram (CMD) to a grid of synthetic CMDs using Bayesian statistics with the Markov-Chain Monte Carlo method to derive age, metallicity, distance and reddening for each cluster (e.g. Kerber et al. 2007, Dias et al. 2014, Maia et al. 2019). Using age and metallicity radial gradients, Dias et al. (2016) inferred that the SMC west halo clusters were being removed from the main body. This was later confirmed by proper motions from Gaia, HST, and VISTA (Zivick et al. 2018; Niederhofer et al. 2018). 

We derive structural parameters (core and tidal radii $r_c, r_t$) based on surface brightness profiles and radial density profile corrected by photometric incompleteness. Completeness is assessed using artificial star tests.
van den Bergh (1991) claimed that LMC cluster diameters increase with increasing distance from the LMC centre. We found this correlation using deprojected distances, but with some dispersion, in particular for Southwest clusters. This is where the warp towards the SMC starts (Choi et al. 2018). 
Also, Miholics et al. (2014) found that the half-mass radii of clusters subject to tidal forces will change rapidly, but the core radii will remain the same. These topics will be discussed in detail in our Paper III (Santos Jr. et al. 2019, in prep.).

Luminosity functions corrected by photometric incompleteness are converted into mass functions using the isochrone that best fits the CMDs of the clusters. VISCACHA data typically reaches down to $0.8-1.0 M_{\odot}$. The mass function slopes are used to detect clusters that are losing low-mass stars (Maia et al. 2019).
We further studied one such case and found that the distribution of main sequence stars is definitely not axisymmetric. This was a strong indication that external forces are destroying this cluster (Paper II: Piatti et al. 2019, submitted).

\section{Summary}

The unique niche explored by the VISCACHA survey and the main science topics where VISCACHA data will play an important role were described. The main techniques, derived cluster parameters and some examples of results were highlighted. The observations are still ongoing to increase the sample and observe more interesting targets in the Magellanic Clouds. The scientific contribution of the VISCACHA project has already started. This is not a public survey, but we understand the legacy value of the project and we intend to release the material in the future, to complement other public surveys.

\end{document}